\documentclass[aps,prd,nofootinbib,notitlepage]{revtex4-1}
\usepackage{enumerate}
\usepackage{dsfont}
\usepackage{amsfonts}
\usepackage{amssymb}
\usepackage[latin1]{inputenc}
\usepackage[T1]{fontenc}
\usepackage{graphicx}
\usepackage{mathtools}
\usepackage{bbold}

\usepackage{soul}
\setstcolor{red}

\usepackage[dvipsnames]{xcolor}

\usepackage[colorlinks=true]{hyperref}
\usepackage{amsthm}

\begin{document}
\title{Berwald spacetimes and very special relativity}

\author{Andrea Fuster}
\email{A.Fuster@tue.nl}
\affiliation{Department of Mathematics and Computer Science, Eindhoven University of Technology, Eindhoven, The Netherlands}

\author{Cornelia Pabst}
\email{pabst@strw.leidenuniv.nl}
\affiliation{Leiden Observatory, Faculty of Science,\\ Leiden University, Leiden, The Netherlands}

\author{Christian Pfeifer}
\email{christian.pfeifer@ut.ee}
\affiliation{Laboratory of Theoretical Physics, Institute of Physics, University of Tartu, W. Ostwaldi 1, 50411 Tartu, Estonia}

\begin{abstract}
In this work we study Berwald spacetimes and their vacuum dynamics, where the latter are based on a Finsler generalization of the Einstein's equations derived from an action on the unit tangent bundle. In particular, we consider a specific class of spacetimes which are non-flat generalizations of the very special relativity (VSR) line element, to which we refer as very general relativity (VGR). We derive necessary and sufficient conditions for the VGR line element to be of Berwald type. We present two novel examples with the corresponding vacuum field equations: a Finslerian generalization of vanishing scalar invariant (VSI) spacetimes in Einstein's gravity as well as the most general homogeneous and isotropic VGR spacetime.
\end{abstract}

\maketitle

%%%%%%%%%%%%%%%%%%%%%%%%%%%%%%%%%%%%%%%%%%%%%%%%%%%%%%%
\section{Introduction}
Already in 1977 Bogoslovsky studied the most general transformations which leave the massless wave equation invariant. These turn out to form the subgroup $DISIM_b(2)$ of the Lorentz group \cite{Bogoslovsky}, the symmetry group on which special relativity is based. The corresponding relativistic kinematics were first studied by Bogoslovsky and Goenner~\cite{Gonner:1997ec,Bogoslovsky:1998wa,Bogoslovsky:1999pp}. Later Cohen and Glashow constructed field theories whose symmetry group is $DISIM_b(2)$ in a framework called  very special relativity (VSR) \cite{Cohen:2006ky}.
These two different approaches to study deformations of Lorentz invariant physics were afterwards  connected by the insight of Gibbons, Gomis and Pope \cite{Gibbons:2007iu}, namely that the symmetries of VSR preserve the line element found by Bogoslovsky
\begin{align}\label{eq:dsVSR}
ds = (\eta_{cd}dx^cdx^d)^\frac{1-b}{2}(n_a dx^a)^b\,,
\end{align}
where $\eta$ is the Minkowski metric, $n=n_adx^a$ is a 1-form with constant components and $b$ is a dimensionless parameter. This line element is a flat Finslerian line element which generalizes the metric line element of special relativity. The deviation from metric geometry is parametrized by the parameter $b$, with the Minkowski spacetime line element recovered in the limit $b=0$.\\

To be able to compare $DISIM_b(2)$ invariant physics with the usual local Lorentz invariant formulation a rigorous discussion of the influence of the Finslerian line element~\eqref{eq:dsVSR} on observables on curved spacetimes is necessary, including dynamical equations which determine the line element. This can be interpreted as the step from very special relativity to a precise notion of \emph{very general relativity} or \emph{general very special relativity} \cite{Gibbons:2007iu,Kouretsis:2008ha}. In what follows we use the abbreviation VGR for both terms. Carrying out this transition from a flat to a curved spacetime geometry one obtains a theory of gravity which is locally $DISIM_b(2)$ invariant. There exist several approaches to this transition in the literature. In \cite{Balan:2011fu}, for example, the Minkowski metric was replaced by a non-flat Lorentzian metric satisfying Einstein's equations and the constant $1$-form $n$ was generalized to a $1$-form $n_a(x) dx^a$ satisfying a Klein-Gordon-like equation. This procedure constructs the curved version of the VSR line element from different dynamical fields on spacetime. Another approach is to consider Finslerian generalizations of Einstein's equations and solve these for curved versions of the line element \eqref{eq:dsVSR}. This was done for example in \cite{Kouretsis:2008ha,Fuster:2015tua}, in the context of cosmology and exact gravitational waves, respectively. In this work we explore this approach in more detail and lift the line element~\eqref{eq:dsVSR} to a Finsler Lagrangian $L$ constructed from a general Lorentzian metric $g$ and a $1$-form $B$. We refer to manifolds equipped with such a Lagrangian as \emph{VGR spacetimes}. The mathematical tools to do so in a precise way are provided by the Finsler spacetime framework proposed in \cite{Pfeifer:2011tk,Pfeifer:2011xi}.\\

The Finsler Lagrangian describing VGR geometries turns out to be a special instance of so-called \mbox{$(\mathcal{A},\mathcal{B})$-Finsler} spacetimes. Their Finsler Lagrangian is a function of two arguments: $\mathcal{A}=g(y,y)$ being the metric length of a vector $y$ and $\mathcal{B}=B(y)$ being a $1$-form acting on a vector. For this class of Finsler spacetimes we compute the geodesic spray explicitly in terms of the Christoffel symbols of metric $g$ and the Levi-Civita covariant derivatives of the $1$-form $B$. This enables us to identify a special class of VGR spacetimes, namely those whose underlying canonical non-linear connection is affine, even though the corresponding length element is of non-trivial Finslerian nature. Finsler spacetimes with this property are called Berwald spacetimes~\cite{Berwald1926,Szilasi2011}.\\

Berwald spacetimes can be regarded as the mildest deviation of a Finslerian geometry from metric geometry. They have recently received attention in the context of modified dispersion relations inspired by quantum gravity \cite{Letizia:2016lew,Lobo:2016lxm}, and in relation to the equivalence principle \cite{Minguzzi2016,Gallego2017}. Another important feature of Berwald spacetimes is that any kind of Finslerian generalization of the Einstein equations should simplify severely for this class due to the relatively simple form of the underlying connection. Among the different suggestions of Finsler generalizations of Einstein's vacuum equations \cite{Chang:2009pa,Kouretsis:2008ha,Stavrinos2014,Asanov,Voicu:2009wi,Bucataru,Rutz,Pfeifer:2011xi,Vacaru:2010fi,Minguzzi:2014fxa} we focus here on the vanishing of the canonical Finsler curvature scalar, derived from the Finslerian geodesic deviation equation~\cite{Rutz}, and on the more involved Finsler spacetime dynamics derived from an action principle on the unit tangent bundle \cite{Pfeifer:2011xi}. The latter two field equations are distinguished because they consider the Finsler function or Lagrangian, i.e.\ a scalar on the tangent bundle of spacetime, as the fundamental dynamical field which determines the geometry of spacetime. Their equivalence has actually been conjectured \cite{Li:2012ty,PhysRevD.90.064049} but a precise mathematical investigation is missing. We compare the two approaches for Berwald spacetimes and prove that they are not equivalent in general. A special class of Finsler spacetimes on which they are equivalent turns out to be VGR Berwald spacetimes built from a non-trivial $1$-form $B$ of vanishing norm with respect to the metric $g$. Finally we discuss two example classes of VGR Berwald spacetimes including their vacuum dynamics. One class is a generalization of the Finsler $pp$-wave spacetimes discussed in \cite{Fuster:2015tua} while the other class is given by the most general homogeneous and isotropic VGR Berwald spacetimes.\\

The paper is organized as follows. In section \ref{sec:FSFST} we recall the notions of Finsler spaces and their generalization to Finsler spacetimes. In section \ref{sec:BerwDyn} we show how the general action-based Finsler gravity field equations simplify for Berwald spacetimes. Afterwards in section \ref{ssec:VGRL} we define VGR spacetimes in general and derive in section \ref{ssec:BerwVGR} conditions for these to be of Berwald type  (Theorem 1). Moreover, we find conditions for $(\mathcal{A},\mathcal{B})$-Finsler spacetimes, and in particular VGR spacetimes, to be not only Berwald, but also have a geometry fully determined by the Levi-Civita connection of the Lorentzian metric involved (Theorem 2). The vacuum dynamics of VGR Berwald spacetimes are discussed in section \ref{sec:BerwVGRDyn}, with special emphasis on the null-VGR case (Theorem 3). In section \ref{sec:ex} we demonstrate our findings with explicit examples, which demonstrate the existence of non-trivial VGR Berwald spacetimes. 

%%%%%%%%%%%%%%%%%%%%%%%%%%%%%%%%%%%%%%%%%%%%%%%%%%%%%%%
\section{Finsler spaces and Finsler spacetimes}\label{sec:FSFST}
Finsler geometry is a well-defined natural extension of Riemannian geometry based on the most general length measure for curves on a manifold \cite{Finsler, Bao, Bucataru}. In the context of relativity, Finsler geometry needs to be adapted such that it extends Lorentzian geometry. However, the generalization of pseudo-Riemannian to pseudo-Finsler geometry is hindered by several issues, which we briefly review in this section. \\

The Finslerian geometry of a manifold $M$ is formulated in terms of tensors on its tangent bundle $TM$. In this work we use the following notation. An element $Y$ of the tangent bundle $TM$ is a vector in some tangent space $T_pM$ at a point $p\in M$. In local coordinates around the point $p=\{x^a\}$, the vector $Y\in T_pM$ can be expressed in the corresponding coordinate basis:
\begin{align}
Y = y^a \frac{\partial}{\partial x^a} \equiv (x,y)\,.
\end{align}
We identify the element $Y\in TM$ with $(x, y)$, which defines the manifold-induced coordinates on the tangent bundle. The canonical coordinate basis of the tangent $T_{(x,y)}TM$ and cotangent $T^*_{(x,y)}TM$ spaces of the tangent bundle will be denoted by:
\begin{align}
T_{(x,y)}TM = \mathrm{span}\bigg\{\frac{\partial}{\partial x^a} = \partial_a, \frac{\partial}{\partial y^a} = \bar{\partial}_a\bigg\},\quad 
T^*_{(x,y)}TM = \mathrm{span}\bigg\{dx^a, dy^a\bigg\}\,.
\end{align}

%++++++++++++++++++++++++++++++++++++++++++++++++%
\subsection{Finsler spaces}\label{ssec:FinslerSpace}
A Finsler space is a smooth manifold $M$ equipped with a smooth real function~$F$
\begin{align}
F: TM\setminus{(x,0)}&\rightarrow \mathbb{R}\\
(x,y)&\mapsto F(x,y)
\end{align}
called the Finsler function, such that:
\begin{itemize}
	\item $F$ is homogeneous of degree one with respect to $y$:
	\begin{align}
	F(x,\lambda y) =\lambda F(x, y)\,,\quad \forall \lambda>0\,;
	\end{align}
	\item $F$ possesses a non-degenerate and positive definite Finsler metric $g^F$ with components:
	\begin{align}
	g^F_{ab} = \frac{1}{2}\bar{\partial}_a \bar{\partial}_b F^2\,;
	\end{align}
	\item $F$ defines the length of curves $\gamma$ on $M$ via the parametrization-invariant length functional:
	\begin{align}\label{lengthofcurve}
	S[\gamma]=\int d\tau\ F(\gamma,\dot{\gamma})\,,\quad \dot{\gamma}=\frac{d\gamma}{d\tau}.
	\end{align}
\end{itemize}
The geometry of $M$ can be derived from $F$ in a similar way as one usually derives the geometry of a metric manifold from the metric tensor. A fundamental ingredient in the construction of the geometry of a Finsler space is the so-called Cartan non-linear connection, defined in terms of its connection coefficients:
\begin{align}
N^a{}_b(x,y) = \frac{1}{4}\bar{\partial}_b \bigg(g^{Fac}\big(y^d\partial_d\bar{\partial}_c F^2 - \partial_c F^2\big)\bigg)\,.
\end{align}
The geodesic equation for curves $\gamma$ on a Finsler space can be written as
\begin{align}
\ddot \gamma^a + N^a{}_b(\gamma,\dot \gamma)\dot \gamma^b = 0\,
\end{align}
where $\ddot \gamma =d^2 \gamma/d\tau^2$ and $\tau$ is an affine parameter. We also define the geodesic spray, which can be used to characterize Berwald spaces, in terms of its coefficients:
\begin{align}
G^a(x,y) = N^a{}_b(x,y) y^b = \frac{1}{2}g^{Fac}\big(y^d\partial_d\bar{\partial}_c F^2 - \partial_c F^2\big)\,.
\end{align}

Finsler spaces reduce to Riemannian manifolds when  $F=\sqrt{g_{ab}(x)y^ay^b} = \sqrt{g(y,y)}$, with $g$ being a Riemannian metric. The Finslerian geodesic equation becomes the geodesic equation on a Riemannian manifold and the non-linear connection coefficients become the Christoffel symbols of the Levi-Civita connection. Finsler geometry is thus a natural generalization of Riemannian geometry.\\

However, the following problem arises when employing Finsler geometry as a generalization of pseudo-Riemannian geometry. As soon as metric $g$ is indefinite, the corresponding Finsler function $F$ is neither smooth nor real on all of $TM\setminus{(x,0)}$ due to the existence of both non-trivial null vectors and vectors with negative metric length. In the context of relativity this is a severe problem since null spacetime directions are interpreted as those along which light propagates. In order to employ Finsler geometry as a generalization of Lorentzian geometry, i.e. of pseudo-Riemannian geometry with a metric of signature $(-,+,+,+)$, we recall the construction of Finsler spacetimes.

%++++++++++++++++++++++++++++++++++++++++++++++++%
\subsection{Finsler spacetimes}
Finsler spacetimes are generalizations of Finsler spaces to manifolds equipped with a Finsler metric of indefinite signature, i.e. a pseudo-Finsler metric. Physically they provide the geometric structures needed to describe observers, their measurement of proper time as well as the motion of massive and massless point particles in a most general way. While several approaches to the construction of Finsler spacetimes have been proposed \cite{Beem,Javaloyes2014}, we consider the framework developed by one of us as most suitable for our purposes. We briefly summarize it here, further details can be found in \cite{Pfeifer:2011tk,Pfeifer:2011xi}.\\

A Finsler spacetime $(M,L)$ is a four-dimensional, connected, Hausdorff, paracompact, smooth manifold~$M$ equipped with a \emph{Finsler Lagrangian}. The latter is a continuous function $L:TM\rightarrow\mathbb{R}$ defined on the tangent bundle with the following properties:
\begin{itemize}
	\item $L$ is smooth on the tangent bundle without the zero section, $TM\setminus{(x,0)}$;\vspace{-6pt}
	\item $L$ is positively homogeneous of real degree $r\ge 2$ with respect to the fibre coordinates of $TM$:
	\begin{equation}\label{eqn:hom}
	L(x,\lambda y)  = \lambda^r L(x,y), \quad \forall \lambda>0\,;
	\end{equation}
	\item $L$ is reversible in the sense:
	\begin{equation}\label{eqn:rev}
	|L(x,-y)|=|L(x,y)|\,;
	\end{equation}
	\item the Hessian $g^L_{ab}$ of $L$ with respect to the fibre coordinates
	\begin{equation}\label{eq:Lmetric}
	g^L_{ab}(x,y) = \frac{1}{2}\bar\partial_a\bar\partial_b L\,
	\end{equation}
	is non-degenerate on $TM\setminus A$, where $A$ has measure zero and does not contain the null set $\{(x,y)\in TM\,|\,L(x,y)=0\}$;
	\item the unit-timelike condition holds, i.e., for all $x\in M$ the set
	\begin{equation}
	\Omega_x=\left\{y\in T_xM\,\Big|\, |L(x,y)|=1\,,\;g^L_{ab}(x,y)\textrm{ has signature }(\epsilon,-\epsilon,-\epsilon,-\epsilon)\,,\, \epsilon=\frac{|L(x,y)|}{L(x,y)}\right\}
	\end{equation}
	contains a non-empty closed connected component $S_x\subset \Omega_x\subset T_xM$.
\end{itemize}
The Finsler function associated with $L$ is then $F(x,y) = |L(x,y)|^{1/r}$ and the corresponding Finsler metric is as usual $g^F_{ab}=\frac{1}{2}\bar\partial_a\bar\partial_bF^2$. This definition of Finsler spacetimes is constructed so as to cover interesting  examples from physics, such as light propagation in area metric geometry and local and linear pre-metric electrodynamics \cite{Rubilar:2007qm,Punzi:2007di,Gurlebeck:2018nme}, which include the bi-metric light-cone structure of birefringent crystals. The most important ingredient in our definition is the use of a $r$-homogeneous function $L$ instead of the $1$-homogeneous function~$F$. This circumvents the problems discussed in the previous section, in particular it avoids the issue about the Finsler function possibly becoming imaginary. Moreover the approach we employ here extends earlier ones, as one sees when the homogeneity degree of $L$ is fixed to $r=2$, since then our definition reduces to the one given by Beem \cite{Beem}. \\

The geometry of Finsler spacetimes is derived from the Lagrangian $L$ and the $L$-metric $g^L$. Similar to Finsler spaces the fundamental ingredient defining the geometry of $M$ is the Cartan non-linear connection, whose connection coefficients are now derived from $L$:
\begin{align}\label{eq:CNL}
N^a{}_b(x,y) = \frac{1}{4}\bar{\partial}_b \bigg(g^{Lac}\big(y^d\partial_d\bar{\partial}_c L - \partial_c L\big)\bigg)\,.
\end{align}
The geodesic equation for curves $\gamma$ on $M$ reads again
\begin{align}
\ddot \gamma^a + N^a{}_b(\gamma,\dot \gamma)\dot \gamma^b = 0
\end{align}
as for Finsler spaces. The geodesic spray coefficients can be expressed as:
\begin{align}\label{eq:geodsprayL}
G^a (x,y) = N^a{}_b(x,y) y^b = \frac{1}{2}g^{Lac}\big(y^d\partial_d\bar{\partial}_c L - \partial_c L\big)\,.
\end{align}

Interestingly \eqref{eq:CNL} is identical to the formulation in terms of the Finsler function $F= |L(x,y)|^{1/r}$ wherever $F$ is differentiable (see \cite{Pfeifer:2011tk} for a proof):
\begin{align}\label{eq:NFNL}
N^a{}_b(x,y) 
&= \frac{1}{4}\bar{\partial}_b \bigg(g^{Lac}\big(y^d\partial_d\bar{\partial}_c L - \partial_c L\big)\bigg)\nonumber\\
&= 	\frac{1}{4}\bar{\partial}_b \bigg(g^{Fac}\big(y^d\partial_d\bar{\partial}_c F^2 - \partial_c F^2\big)\bigg)\,.
\end{align}

As an aside we note that the connection coefficients can identically be derived from any power $m$ of $L$, i.e., $N^a{}_b[L] = N^a{}_b[L^m]$, so in particular, as displayed above, $N^a{}_b[L] = N^a{}_b[|L|^\frac{2}{r}] = N^a{}_b[F^2]$, taking the norm does not make a difference. The non-linear connection defines so-called horizontal derivative operators, the horizontal basis of tangent spaces of the tangent bundle
\begin{align}\label{eq:horr}
\delta_a = \partial_a - N^b{}_{a}(x,y)\partial_b\,,
\end{align}
whose special property is that it transforms as a tensor under a manifold-induced coordinate change of the base manifold, i.e., $(x,y) \mapsto ( \tilde{x}(x), \tilde y(x,y))$ with $\tilde y^a=y^b\partial_b \tilde x^a$ implies $\tilde \delta_a = \tilde \partial_a x^b \delta_b$. \\

Dynamics of Finsler spacetimes can be obtained in terms of the curvature derived from the Cartan non-linear connection which drives the gravitational tidal forces, or in mathematical terms, the geodesic deviation:
\begin{align}
R^a{}_{bc}(x,y) = [\delta_b,\delta_c]^a = \delta_c N^{a}{}_b(x,y) - \delta_b N^{a}{}_c(x,y).
\end{align}
From here we can construct the canonical Finsler curvature scalar
\begin{align}\label{eq:FCurvScal}
\mathcal{R}(x,y) = R^a{}_{ac}(x,y)y^c \equiv \mathcal{R}_{ab}(x,y)y^ay^b\,,
\end{align}
where $\mathcal{R}_{ab}(x,y)$ is the Finsler Ricci tensor proposed by Akbar-Zadeh \cite{Akbar-Zadeh1988}. This scalar is the building block for the dynamical field equations of the Finsler Lagrangian, which we derive for Berwald spacetimes in the next section.

%%%%%%%%%%%%%%%%%%%%%%%%%%%%%%%%%%%%%%%%%%%%%%%%%%%%%%%
\section{Finsler gravity vacuum dynamics for Berwald spacetimes}\label{sec:BerwDyn}
Berwald spacetimes are minimal Finslerian extensions of pseudo-Riemannian spacetimes. They are equipped with a non-metric Finsler Lagragian but their non-linear connection is still an affine connection \cite{Szilasi2011}. An equivalent characterization is that their geodesic spray is quadratic in the velocities $y$. The precise mathematical condition for a Finsler spacetime to be of Berwald type is:
\begin{align}\label{eq:BerwaldCond}
	\bar \partial_d \bar \partial_c \bar \partial_b G^a(x,y) = 0\,.
\end{align}
This condition implies the following:
\begin{align}\label{eq:GeodSprayBerw}
	G^a(x,y) = \mathcal{G}^a{}_{ij}(x)y^iy^j \textrm{ and } N^{a}{}_b = \mathcal{G}^a{}_{(bj)}(x)y^j\,,
\end{align}
where $\mathcal{G}^a{}_{ij}(x)$ are coefficients of an affine connection. In other words the geometric structure of Berwald spacetimes is minimally more general than that of metric spacetimes. In fact there exist Berwald spacetimes whose geometry is identical to metric spacetime geometry. For these the only difference to their metric counterparts is the length measure employed for curves (an example is given in section \ref{sec:ex}). Berwald spacetimes are the most conservative Finslerian extension of metric spacetime geometry and thus plausible spacetime candidates for extended theories of gravity based on generalized geometries.\\

Our goal now is to obtain the vacuum dynamics for Berwald spacetimes, generalizing Einstein's equations. In what follows we employ the action-based vacuum field equation for general Finsler spacetimes developed in \cite{Pfeifer:2011xi}
\begin{align}\label{eq:Fgrav}
\frac{r}{2} L g^{L ab}\bar{\partial}_a\bar \partial_b \mathcal R - \frac{2(2r-1)}{(r-1)}\mathcal{R} + r L g^{L ab} \left( \nabla^B_a S_b + \bar \partial_a (y^q \nabla^B_q S_b) \right) = 0\,. 
\end{align}
where $\nabla^B$ is the Berwald covariant derivative and $S$ is the Landsberg tensor (see appendix \ref{app:def}). Note that this is a \emph{scalar} equation on the tangent bundle, determined by the Finsler Lagragian $L$ and its derived objects. In the case of a metric Finsler Lagrangian, $L= g_{ab}(x)y^ay^b$, the Finsler spacetime vacuum equation is equivalent to Einstein's vacuum equations. The Landsberg Tensor $S$ vanishes identically for Berwald spacetimes, which simplifies the field equation considerably:
\begin{align}\label{eq:FgravBerw}
	\frac{r}{2} L g^{L ab}\bar{\partial}_a\bar{\partial}_b \mathcal R - \frac{2(2r-1)}{(r-1)}\mathcal{R} = 0\,.
\end{align}
Moreover, the curvature scalar \eqref{eq:FCurvScal} can be derived from the geodesic spray \eqref{eq:GeodSprayBerw} as
\begin{align}\label{eq:RicciBerw}
	\mathcal{R}(x,y)
	&= y^c \delta_a N^a{}_c - y^c \delta_c N^a{}_a 
	=y^jy^c\left( \partial_a\mathcal{G}^a{}_{(cj)} -\partial_c \mathcal{G}^a{}_{(aj)} +  \mathcal{G}^a{}_{(ar)} \mathcal{G}^r{}_{(cj)} - \mathcal{G}^a{}_{(cr)} \mathcal{G}^r{}_{(aj)}\right)\nonumber\\
	&\equiv y^j y^c \mathcal{R}_{jc}(x)\,
\end{align}
where we define the Berwald Ricci tensor $\mathcal{R}_{jc}(x)$, which is just the Finsler Ricci tensor for Berwald spacetimes. Note that, in contrast to the general case, it does not depend on the velocities $y$. Employing this result we can reformulate the Finsler spacetime dynamics on Berwald spacetimes \eqref{eq:FgravBerw} to
\begin{align}\label{eq:FGravBerwald}
	\bigg( L g^{L ab} - \frac{(2r-1)}{(r-1)} y^a y^b \bigg) \mathcal{R}_{ab}  = 0\,,
\end{align}
which we call \emph{Berwald gravity equation} from now on. This equation is the generalization of Einstein's vacuum equations to Berwald spacetimes and it is derived here for the first time.\\

We finish the section by commenting on the relation to earlier work on Finsler gravity vacuum dynamics. In \cite{Li:2012ty} it was suggested that the field equation \eqref{eq:Fgrav} is equivalent to the one suggested by Rutz, $\mathcal{R}_{ab}(x,y)y^ay^b=0$ \cite{Rutz}. Berwald spacetimes demonstrate that this is only the case for very special Finsler spacetimes. On Berwald spacetimes Rutz's equation reduces to $\mathcal{R}_{ab}(x)y^ay^b = 0$, which implies $\mathcal{R}_{ab}(x)=0$. Hence Rutz's equation implies that \eqref{eq:FGravBerwald} (which is \eqref{eq:Fgrav} for Berwald spacetimes) holds. The converse is however not true. Assume that \eqref{eq:FGravBerwald} holds; this does not imply $\mathcal{R}_{ab}(x)y^ay^b = 0$, unless $g^{Lab}\mathcal{R}_{ab}(x)=0$ is also satisfied. Thus Berwald spacetimes show that field equation \eqref{eq:Fgrav} is in general not equivalent to $\mathcal{R}_{ab}(x,y)y^ay^b = 0$, although there may exist very special Finsler spacetimes for which this is the case.

%%%%%%%%%%%%%%%%%%%%%%%%%%%%%%%%%%%%%%%%%%%%%%%%%%%%%%%
\section{Very general relativity}\label{sec:VGR}
We now turn to the study of VGR spacetimes which are of Berwald type. In order to do so we compute the VGR geodesic spray and derive a necessary and sufficient condition for the spacetime to be Berwald (Theorem 1). Moreover we find a sufficient condition such that a $(\mathcal{A},\mathcal{B})$-Finsler spacetime, and in particular a VGR spacetime, is not only Berwald but its geometry is fully determined by the Levi-Civita connection of the Lorentzian metric involved (Theorem~2). And we show that for null-VGR Berwald spacetimes the vanishing of the Berwald Ricci tensor is equivalent to the action-based Finsler dynamics (Theorem 3). Last we demonstrate our findings by displaying examples of VGR Berwald spacetimes.

%++++++++++++++++++++++++++++++++++++++++++++++++%
\subsection{VGR spacetimes}\label{ssec:VGRL}
We recall the length element of very special relativity introduced in Eq.~(\ref{eq:dsVSR}):
\begin{align}\label{eq:dsVSR2}
ds = (\eta_{cd}dx^cdx^d)^\frac{1-b}{2}(n_a dx^a)^b\,
\end{align}
In order to lift (\ref{eq:dsVSR2}) to a length element of a curved spacetime we replace the Minkowski metric $\eta$ with a general Lorentzian metric $g$, and the constant components $1$-form $n$ with a general $1$-form $B$. We then obtain the (Finsler function representation of the) length element:
\begin{align}\label{dsVGR}
	\tilde F = (g_{cd}(x)y^c y^d)^\frac{1-b}{2}(B_a(x) y^a)^b = g(y,y)^\frac{1-b}{2}B(y)^b\,.
\end{align}
As explained in section \ref{ssec:FinslerSpace} such Finsler functions are problematic for indefinite metrics $g$, since they are not \mbox{differentiable} on the null structure of the theory and may also become imaginary. In order to obtain a \mbox{differentiable}, real Finsler Lagragian which properly defines VGR spacetimes we take an appropriate power of the Finsler function as Lagrangian 
\begin{align}\label{eq:LVGR}
	L = g(y,y)B(y)^n\,
\end{align}
where $n = \frac{2 b}{1-b}$. Wherever both $\tilde F$ and $L$ are differentiable they define the same geometry of spacetime as explained around Eq.~\eqref{eq:NFNL}, since $L$ is the $\frac{2}{1-b}$ power of $\tilde F$. We call manifolds $M$ equipped with a Finsler Lagragian $L$ of the type~\eqref{eq:LVGR}~\emph{VGR spacetimes}. The VGR Finsler Lagragian remains problematic where $B(y)$ vanishes, since the corresponding L-metric \eqref{eq:Lmetric} is not invertible\footnote{The $L$-metric and its inverse are given in section \ref{sec:BerwVGRDyn}.}. The influence of this set depends on the causal character of the $1$-form $B$ with respect to the metric $g$, and may lead to further constraints on the form of the Lagragian. We do not tackle this issue here and thus all derivations below hold everywhere except on this set. \\

The Finsler function which defines the length of curves on VGR spacetimes is $F = |L|^\frac{1}{n+2}$. The length measure on VGR spacetimes is thus given by, recall Eq.~(\ref{lengthofcurve}):
\begin{align}
	S[\gamma] = \int d\tau\ |g(\dot \gamma, \dot{\gamma}) B(\dot\gamma)^n|^{\frac{1}{n+2}}\,.
\end{align}
In order to study VGR spacetimes we need to derive the Finsler geometric objects of $L$, with the geodesic spray being the fundamental ingredient. First, we observe that the VGR Lagragian~\eqref{eq:LVGR} is a particular function of the variables $\mathcal{A}=g(y,y)$ and $\mathcal{B} = B(y)$:
\begin{align} \label{VGR-AB}
	L = \mathcal{A}\ \mathcal{B}^n\,.
\end{align}
We derive next the geodesic spray of VGR spacetimes using Eq.~\eqref{eq:Ga}, which is the geodesic spray for general Finsler Lagrangians of the type $L=L(\mathcal{A},\mathcal{{B}})$. For VGR spacetimes we have
\begin{align}
\partial_{\mathcal{A}} L = \mathcal{B}^n,&\quad \partial_{\mathcal{B}} L = n \mathcal{A}\ \mathcal{B}^{n-1},\\
\partial_{\mathcal{A}} \partial_{\mathcal{A}} L = 0,\quad
\partial_{\mathcal{A}} \partial_{\mathcal{B}} L &= n \mathcal{B}^{n-1},\quad
\partial_{\mathcal{B}} \partial_{\mathcal{B}} L = n (n-1)\mathcal{A}\ \mathcal{B}^{n-2}
\end{align}
and the desired geodesic spray is thus 
\begin{align}\label{eq:VGRG}
2 G^a &= 2 \Gamma^a{}_{bc}y^by^c + n \frac{\mathcal{A}}{\mathcal{B}}y^b g^{ac} (\nabla_b B_c - \nabla_c B_b)\nonumber\\
&+ \frac{n(2 \mathcal{B}\ y^a - \mathcal{A}\ B^a)}{2( 1 + n)\mathcal{B}^3 - n g(B,B) \mathcal{A}\ \mathcal{B}}\bigg(n \mathcal{A} y^b B^c (\nabla_c B_b - \nabla_b B_c) + 2 \mathcal{B} y^b y^c \nabla_b B_c\bigg)\,
\end{align}
where $\Gamma^a{}_{bc}$ are the Christoffel symbols of the Lorentzian metric $g$, $\nabla$ denotes the Levi-Civita covariant derivative and, by abuse of notation, the norm of the $1$-form $B$ with respect to the metric is denoted by $g(B,B) = g^{ab}B_aB_b$.
From the expression above it is clear that the parameter $n$ and the Levi-Civita covariant derivatives of $B$ define the deviation of VGR spacetime geometry from metric spacetime geometry, recovered for $n=0$. All further geometric objects, such as the non-linear connection and the curvature, can be derived from the geodesic spray.

%++++++++++++++++++++++++++++++++++++++++++++++++%
\subsection{VGR Berwald spacetimes}\label{ssec:BerwVGR}
We identify next a class of VGR spacetimes which can be interpreted as a minimal deviation from metric geometry, namely \emph{VGR Berwald spacetimes}. These should yield a geodesic spray quadratic in the tangent directions~$y$, recall section \ref{sec:BerwDyn}. It can be seen from Eq.~\eqref{eq:VGRG} that the Levi-Civita covariant derivative of the $1$-form $B$ and the parameter $n$ determine whether the VGR spacetime is of Berwald type or not. In order to derive precise conditions we split the covariant derivative into a symmetric and an antisymmetric part in the following form
\begin{align}
\nabla_a B_b = P B_a B_b + Q D_{ab} + g(B,B) E_{ab}\,
\end{align} 
where $D_{ab}$ and $E_{ab}$ are the components of symmetric and antisymmetric $(0,2)$-tensors and $P,Q$ are functions on spacetime we seek to determine. Plugging this ansatz into the geodesic spray \eqref{eq:VGRG} yields:
\begin{subequations}
\begin{align}
2 G^a = 2 \Gamma^a{}_{bc}(x)y^b y^c 
&+ 2 n R g(B,B) \mathcal{A}\bigg( \frac{E_b{}^ay^b}{\mathcal{B}} - \frac{n E_{bc}y^bB^c(2 \mathcal{B}y^a - \mathcal{A}B^a)}{2(1+n)\mathcal{B}^3 - n g(B,B)\mathcal{A}\mathcal{B}} \bigg)\label{eq:VGRBerwCond1}\\
&+ \frac{2 n (P \mathcal{B}^2+Q y^by^cD_{bc})}{2(1+n)\mathcal{B}^2 - n g(B,B)\mathcal{A}}(2 \mathcal{B}y^a - \mathcal{A}B^a)\,.\label{eq:VGRBerwCond2}
\end{align}
\end{subequations}
In order for these to be quadratic in the velocities $y$ the following conditions have to be satisfied. On the one hand, the fraction in \eqref{eq:VGRBerwCond2} must be independent of $y$ since the multiplying factor is already quadratic in $y$. The only way to achieve this is to set $P=2(1+n) C(x)$ and $Q D_{bc} = - C(x) n g(B,B) g_{b
c}$. This is derived from the fact that equations of the type
\begin{align}
\frac{T_{ab}(x)y^ay^b}{Z_{cd}(x)y^cy^d} = W(x)
\end{align}
can only hold if $T_{(ab)} = W(x) Z_{(ab)}$. In our case the tensors $T$ and $Z$ are already symmetric, so $T_{ab} = W(x) Z_{ab}$. This yields the conditions on $P,Q$ and $D_{bc}$ listed above.\\

On the other hand, contracting the bracket in \eqref{eq:VGRBerwCond1} with $B_a$ yields the necessary condition that the resulting scalar multiplied by $\mathcal{A}$ must be quadratic in $y$
\begin{align}
\frac{\mathcal{A}\mathcal{B} E_{bc}y^b B^c}{2(1+n)\mathcal{B}^2-ng(B,B)\mathcal{A}} = V^a{}_{bc}(x)y^by^c B_a
\end{align}
which can be solved for:
\begin{align}
E_{bc}y^b B^c =\frac{1}{\mathcal{A}\mathcal{B}} V^a{}_{bc}(x)y^by^c B_a \left( 2(1+n)\mathcal{B}^2-ng(B,B)\mathcal{A}\right).
\end{align}
However, this equation has no solutions for all $y$ except $E_{bc}=0 = V^a{}_{bc}$ since there exists no $V^a{}_{bc}$ such that the polynomials on the right-hand side combine to a first order monomial in $y$. We summarize the result in the following theorem.\\

\noindent \textbf{Theorem 1.} \emph{Let $(M,L)$ be a VGR spacetime, i.e. $L = g(y,y)B(y)^n$. A VGR spacetime is of Berwald type if and only if there exists a function $C(x)$ such that the $1$-form $B$ satisfies: 
\begin{equation}\label{eq:Thm1cond}
\nabla_aB_b = C(x) \left( 2(1+n)B_a B_b - n g(B,B) g_{ab}\right).
\end{equation}
The geodesic spray of such a spacetime is quadratic in the directions $y$ and reads as follows:
\begin{align}\label{eq:geodsprayvgrberwald}
 G^a = \Gamma^a{}_{bc}y^by^c  + n C(x)  (2 \mathcal{B} y^a - \mathcal{A}B^a)\,.
\end{align}
For the specific class of null-VGR spacetimes, i.e.\ VGR spacetimes for which $g(B,B)=0$, the Berwald condition (\ref{eq:Thm1cond}) simplifies to: 
\begin{equation}\label{eq:Thm1condnull}
\nabla_aB_b = C(x) 2(1+n)B_a B_b\,.
\end{equation}}
This theorem identifies a necessary and sufficient condition for a VGR spacetime, with general metric $g$ and 1-form $B$, to be of Berwald type. An earlier sufficient condition (in terms of the existence of a special vielbein) for Finsler spacetimes of the form (\ref{dsVGR}) and a Lorentzian metric $g$ belonging to Kundt's class \cite{Kundt1961,stephani_kramer_maccallum_hoenselaers_herlt_2003,FusterPerez:2007zz} can be found in \cite{Gomez-Lobo:2016qik}. A simpler class of VGR Berwald spacetimes can be found from the next theorem for general $(\mathcal{A},\mathcal{B})$-Finsler spacetimes, which we prove in Appendix~\ref{app:geodSprayL(A,B)}.\\

\noindent \textbf{Theorem 2.} \emph{Let $(M,L)$ be a $(\mathcal{A},\mathcal{B})$-Finsler spacetime, i.e. $L=L(\mathcal{A},\mathcal{B})$ with $\mathcal{A}=g(y,y)$ and $\mathcal{B}=B(y)$. A \mbox{sufficient} condition for $(M,L)$ to be a Berwald spacetime is that $B$ is covariantly constant with respect to the Lorentzian \mbox{metric}~$g$. The geodesic spray of $(M,L)$ is then given by $G^a = \Gamma^a{}_{bc}y^by^c$, where $\Gamma^a{}_{bc}$ are the Christoffel symbols of the metric.} \\

Note that the existence of such a theorem, which we proved now, was conjectured in \cite{Schreck:2015seb} on the basis of numerical calculations. Moreover similar results for Randers spaces have been known for a long time \cite{matsumoto1974,Hashiguchi1975,Shibata1977,Kikuchi1979}, and more recently for Finsler $b$-spaces in the context of Lorentz-violating extensions of the Standard Model \cite{Kostelecky:2011qz}. For VGR spacetimes Theorem 2 implies that for a covariantly constant (c.c.\,from now on) $1$-form $B$, i.e. $\nabla_aB_b=0$ and so $C(x)=0$, the geometry of the VGR Berwald spacetime is fully determined by the Christoffel symbols of the Lorentzian metric $g$. Therefore, the geometry does not depend on the exponent parameter $n$ and all VGR Berwald spacetimes $L_n = g(y,y)B(y)^n$ yield the same geometry regardless of $n$. The difference between families of VGR Berwald spacetimes $(M,L_n)$ and the metric spacetimes $(M,g)$ of general relativity becomes only apparent when employing the $L$-metric, and in the identification of normalized timelike vectors. The case of a c.c.\,1-form $B$ may be seen as the mildest deviation of a VGR model from general relativity.

%++++++++++++++++++++++++++++++++++++++++++++++++%
\subsection{Dynamics of VGR Berwald spacetimes}\label{sec:BerwVGRDyn}
We consider now the vacuum dynamics of VGR Berwald spacetimes. These can be derived from the Berwald gravity vacuum Eq.~\eqref{eq:FGravBerwald}, for which we need to compute the (inverse) $L$-metric for VGR spacetimes of the type \eqref{VGR-AB}. Note that the homogeneity degree $r$ of $L$ is related to the exponent parameter n as $r=n+2$. With the help of the formulas in Appendix~\ref{app:geodSprayL(A,B)} we calculate
\begin{align}
	g^L_{ab} = \mathcal{B}^n g_{ab}  + n \mathcal{B}^{n-1} (B_a y_b + B_b y_a) + \frac{n}{2}(n-1) \mathcal{A} \mathcal{B}^{n-2} B_a B_b 
\end{align}
and
\begin{align}
	L g^{Lab} = \mathcal{A} g^{ab}
	&- \frac{2 n \mathcal{B} \mathcal{A}}{2(1+n)\mathcal{B}^2 -n g(B,B)\mathcal{A}}(B^a y^b + B^b y^a) \nonumber \\
	&+ \frac{2 n^2 g(B,B)\mathcal{A}}{(1+n)\left( 2(1+n)\mathcal{B}^2 -n g(B,B)\mathcal{A} \right)} y^ay^b\nonumber \\
	&+ \frac{n \mathcal{A}^2}{2(1+n)\mathcal{B}^2 -n g(B,B)\mathcal{A}} B^a B^b\,.
\end{align}
The Berwald vacuum gravity equation for VGR spacetimes thus becomes
\begin{align}
	\Bigg(\mathcal{A} g^{ab}
	&- \frac{2 n \mathcal{B} \mathcal{A}}{2(1+n)\mathcal{B}^2 -n g(B,B)\mathcal{A}}(B^a y^b + B^b y^a)\nonumber\\
	&+ \bigg(\frac{2 n^2 g(B,B)\mathcal{A}}{(1+n)(2(1+n)\mathcal{B}^2 -n g(B,B)\mathcal{A})} - \frac{2n + 3}{n+1}\bigg) y^ay^b\nonumber\\
	&+ \frac{n \mathcal{A}^2}{2(1+n)\mathcal{B}^2 -n g(B,B)\mathcal{A}} B^a B^b\Bigg)\mathcal{R}_{ab}=0\,.
\end{align}
Recall that Berwald spacetimes have the property $\mathcal{R}_{ab}=\mathcal{R}_{ab}(x)$. Next, we multiply the expression by the denominator \\$(1+n)(2(1+n)\mathcal{B}^2 -n g(B,B)\mathcal{A})$ and obtain a fourth-order polynomial in $y$:
\begin{align}\label{eq:VGRBerwaldFinsGrav}
	\mathfrak{G}(x,y)
	=&\bigg((2(1+n)\mathcal{B}^2 -n g(B,B)\mathcal{A})\left(\mathcal{A} g^{ab} - \tfrac{2n+3}{n+1}y^ay^b\right)\nonumber\\
	&- 2 n \mathcal{B}\mathcal{A}(B^a y^b + B^b y^a) + 2 n^2 g(B,B)\mathcal{A} y^a y^b + n \mathcal{A}^2 B^a B^b\bigg)\mathcal{R}_{ab}=0\,.
\end{align}
Taking a fourth-order derivative with respect to $y$ yields a purely tensorial equation on spacetime which determines the dynamics of the $1$-form $B$ and the Lorentzian metric $g$:
\begin{align}\label{eq:VGRBerwaldTensorEq}
	\bar\partial_a\bar\partial_b\bar\partial_c\bar\partial_d\mathfrak{G}(x,y) = \mathfrak{G}_{abcd}(x) = 0\,.
\end{align}
We now analyse the above equation for null-VGR Berwald spacetimes, i.e.\ those for which $g(B,B)=0$. In this case Eq.~\eqref{eq:VGRBerwaldFinsGrav} reduces to
\begin{align}\label{eq:VGRBerwaldFinsGravnull}
\mathfrak{G}(x,y)
=&\bigg(2(1+n)\mathcal{B}^2\left(\mathcal{A} g^{ab} - \tfrac{2n+3}{n+1}y^ay^b\right)- 4 n \mathcal{B}\mathcal{A}B^ay^b + n \mathcal{A}^2 B^a B^b\bigg)\mathcal{R}_{ab}=0\,.
\end{align}
Studying the corresponding tensor Eq.~\eqref{eq:VGRBerwaldTensorEq} yields the integrability condition $\mathcal{R}_{ab}B^a B^b = 0$ (from \mbox{$g^{ab}g^{cd}\mathfrak{G}_{abcd}=0$}). Evaluating $\mathfrak{G}(x,y)$ on this condition yields: 
\begin{align}\label{eq:VGRBerwaldFinsGravnull2}
\mathfrak{G}(x,y)|_{\mathcal{R}_{ab}B^a B^b = 0} \equiv \mathfrak{G}_R(x,y)=\bigg(2(1+n)\mathcal{B}\left(\mathcal{A} g^{ab} - \tfrac{2n+3}{n+1}y^ay^b\right)- 4 n \mathcal{A}B^ay^b\bigg)\mathcal{R}_{ab}=0\,.
\end{align}
The third derivative of this equation with respect to $y$ and its contractions yield additional integrability conditions:
\begin{align}
	y^a g^{bc}\bar{\partial}_a\bar{\partial}_b\bar{\partial}_c\mathfrak{G}_R(x,y)=0 \Leftrightarrow \mathcal{R}_{ab}y^aB^b = \mathcal{R}^a{}_a \mathcal{B} \frac{3+4n}{2(3+8n)}\\
	y^a y^{b}B^c\bar{\partial}_a\bar{\partial}_b\bar{\partial}_c\mathfrak{G}_R(x,y)=0 \Leftrightarrow \mathcal{R}_{ab}y^aB^b = \mathcal{R}^a{}_a \mathcal{B} \frac{1+n}{3+4n}\,.
\end{align}
Thus either $n = -\frac{3}{2}$ or $\mathcal{R}^a{}_a=0$ and $\mathcal{R}_{ab}y^b B^a =0$. The latter condition reduces the field Eq.~\eqref{eq:VGRBerwaldFinsGravnull2} to $\mathcal{R}_{ab}y^ay^b = 0$, which for Berwald spacetimes implies $\mathcal{R}_{ab}=0$. Alternatively, fixing $n$ yields the expression $6 R_{ab}y^b B^a = \mathcal{R}^a{}_a \mathcal{B}$ and reduces the field Eq.~\eqref{eq:VGRBerwaldFinsGravnull2} to:
\begin{align}
	\mathcal{A} \mathcal{R}^a{}_a - 3 \mathcal{R}_{ab}y^ay^b = 0\,.
\end{align}
Another second derivative with respect to the directions $y$ and contraction of the resulting equation with the spacetime metric components $g^{ab}$ enforces $\mathcal{R}^a{}_a=0$ and hence, also in this case, $\mathcal{R}_{ab}y^ay^b = 0$ implying again $\mathcal{R}_{ab}=0$. We summarize the discussion in the following theorem.\\

\noindent \textbf{Theorem 3.} \emph{Let $(M,L)$ be a null-VGR Berwald spacetime, i.e.\ $L = g(y,y)B(y)^n$ with a $1$-form $B$ satisfying $g(B,B)=0$ and $\nabla_aB_b = C(x)2 (1+n) B_a B_b$. The Berwald gravity vacuum equation  
\begin{align}\label{eq:non-nullVGRdynamics}
\bigg( L g^{L ab} - \frac{(2r-1)}{(r-1)} y^a y^b \bigg) \mathcal{R}_{ab}  = 0\,
\end{align}
is equivalent to the vanishing of the Berwald Ricci tensor:
\begin{align}\label{eq:nullVGRdynamics}
	\mathcal{R}_{ab}=0\,.
\end{align}}This result also implies that for null-VGR Berwald spacetimes Rutz's equation $\mathcal{R}_{ab}y^ay^b = 0$ is equivalent to \eqref{eq:non-nullVGRdynamics}, which is in general not true as we discussed below Eq.~\eqref{eq:FgravBerw}. In the case of a non-null VGR Berwald spacetime the dynamics are governed by Eq.~\eqref{eq:non-nullVGRdynamics}, with $\mathcal{R}_{ab}=0$ being a sufficient but not necessary condition. The field equation in this case will be analyzed in more detail in future work. 

%%%%%%%%%%%%%%%%%%%%%%%%%%%%%%%%%%%%%%%%%%%%%%%%%%%%%%%%
\subsection{Examples of VGR Berwald spacetimes}\label{sec:ex}
We present next two novel explicit examples, which demonstrate the existence of non-trivial (null and non-null) VGR Berwald spacetimes. They represent a VGR generalization of vanishing scalar invariant (VSI) spacetimes and the most general homogeneous and isotropic VGR Berwald spacetime. We also discuss their corresponding vacuum dynamics.

%+++++++++++++++++++++++++++++++++++++++++++++++++++++%
\subsubsection{Null-VGR Berwald spacetimes: Finsler VSI spacetimes}  
We present a generalization of the Finsler $pp$-waves in \cite{Fuster:2015tua}, given by the following VGR Lagrangian (recall \mbox{$L=g(y,y)B(y)^n$})
\begin{align}\label{eq:LVGR-VSI}
L = \left(-2(y^{u}y^{v}+\big[ \Phi(u,x^i)+v\,\tilde{\Phi}(u,x^i) \big]\;(y^{u})^2 +W_1(u,x^i)\;y^{u}y^1 +W_2(u,x^i)\;y^{u}y^2   ) + (y^1)^2 + (y^2)^2\right)\left( y^{u}\right)^n\,
\end{align}
expressed in coordinates $(u,v,x^1,x^2;y^u,y^v,y^1,y^2)$, where $u=(1/\sqrt{2})(t-x^3)$ and $v=(1/\sqrt{2})(t+x^3)$ are light-cone coordinates, and $\Phi$, $\tilde{\Phi}$, $W_1$ and $W_2$ are real functions. The Lorentzian metric $g$ is given by: 
\begin{align} \label{VSIsubclass}
ds^2=-2du\left(dv + \big[ \Phi(u,x^i)+v\,\tilde{\Phi}(u,x^i) \big]\, du +W_1(u,x^i)\,dx^1+W_2(u,x^i)\,dx^2\right) +(dx^1)^2 + (dx^2)^2\,  . 
\end{align}
This metric is of Kundt type and belongs to the class of VSI spacetimes \cite{Pravda:2002us}, and in particular to the subclass where metric functions $W_1$ and $W_2$ do not depend on coordinate $v$. This is an exact vacuum solution of Einstein's equations for certain functional dependencies of $\Phi$ and $\tilde{\Phi}$ in terms of $W_i$. In the case $\tilde{\Phi}=0$ it reduces to the gyratonic $pp$-wave metric~\cite{gyraton,Maluf2018}, and to the usual $pp$-waves for $\tilde{\Phi}=W_1=W_2=0$. \\  

Note that \eqref{eq:LVGR-VSI} reduces to the (Lorentzian) Lagrangian $L=g(y,y)$ induced by the VSI metric \eqref{VSIsubclass} in the case $n=0$, and to the (Finsler) Lagrangian induced by the Finsler $pp$-waves in \cite{Fuster:2015tua} for vanishing metric functions $\tilde{\Phi}=W_1=W_2=0$. The case $\tilde{\Phi}=0$ provides a Finsler version of the gyratonic $pp$-wave metric, presented here for the first time. The $1$-form defining the VGR spacetime in consideration is $B = du$. It is null with respect to the Lorentzian metric $g$: $g( B,B) =0$ and its covariant derivative has one non-zero component $\nabla_u B_u=\tilde{\Phi}(u,x^i)$. Thus $B$ satisfies the VGR Berwald condition stated in Eq.~\eqref{eq:Thm1condnull} of Theorem 1 with $C(x)=\tilde{\Phi}(u,x^i)/(2(n+1))$, which reduces to $C(x)=0$ for the Finsler gyratonic $pp$-waves. Our result is completely consistent with Theorem 2 in \cite{Gomez-Lobo:2016qik}, which states that a Finsler spacetime of the form (\ref{dsVGR}) and a Lorentzian metric $g$ of the Kundt type with $W_{i,v}=0$ is always Berwald. \\

Calculating the geodesic spray, using Eq.~\eqref{eq:geodsprayvgrberwald}, yields 
\begin{align}
	G^u&= - \frac{(y^u)^2}{1+n}\tilde{\Phi}\\
	G^v&= \frac{(y^u)^2}{(1+n)} \Big( (1+n)\big( \tilde\Phi (W_1^2 + W_2^2) + W_1 \partial_u W_1 + W_2 \partial_u W_2 + (\partial_u - W_1\partial_{x^1} - W_2 \partial_{x^2})( \Phi + v \tilde\Phi) \big) + (2+n)\tilde\Phi (\Phi + v \tilde \Phi) \Big)\nonumber\\
	   &- \frac{y^u\; y^1}{1+n} \Big( (1+n) \big( W_2 (\partial_{x^2}W_1 - \partial_{x^1}W_2 ) - 2 \partial_{x^1}(\Phi + v \tilde \Phi)\big) + n W_1 \tilde \Phi \Big) + 2 y^u y^v \tilde \Phi \nonumber + (y^1)^2 \Big( \tfrac{n}{2(1+n)} \tilde \Phi + \partial_{x^1} W_1  \Big)\\
	   &+ \frac{y^u\; y^2}{1+n} \Big( (1+n) \big( W_1 (\partial_{x^2}W_1 - \partial_{x^1}W_2 ) + 4 \partial_{x^2}(\Phi + v \tilde \Phi)\big) - n W_2 \tilde \Phi \Big)+ (y^2)^2 \Big( \tfrac{n}{2(1+n)} \tilde \Phi + \partial_{x^2} W_2 \Big)
	   + y^1\; y^2 \Big(\partial_{x^2}W_1 + \partial_{x^1}W_2\Big)\\
	G^1&=(y^u)^2\Big(\partial_{x^1}(\Phi + v \tilde \Phi) - W_1 \tilde\Phi - \partial_u W_1 \Big)  + \frac{n}{n+1}y^u y^1 \tilde \Phi + y^u\; y^2 \Big(\partial_{x^1}W_2 - \partial_{x^2}W_1\Big)\\
	G^2&=(y^u)^2\Big(\partial_{x^2}(\Phi + v \tilde \Phi) - W_2 \tilde\Phi - \partial_u W_2 \Big) + \frac{n}{n+1}y^u y^2 \tilde \Phi + y^u\; y^1 \Big(\partial_{x^2}W_1 - \partial_{x^1} W_2\Big)\,
\end{align}
where we suppressed the arguments of the metric functions for the sake of readability. In the case $\tilde \Phi = 0$ the 1-form $B$ is c.c.\ (since $C(x)=0$) and the geodesic spray reduces to $G^a=\Gamma^a{}_{bc}y^by^c$ (recall section \ref{ssec:BerwVGR}), with $\Gamma^a{}_{bc}$ being the Christoffel symbols of the gyratonic $pp$-wave metric. The Berwald Ricci tensor can now be computed using Eq.~(\ref{eq:RicciBerw}), and the field dynamics $\mathcal{R}_{ab}(x)=0$ (Theorem 3) become:
\begin{align}
	0 =\mathcal{R}_{uu}&= - \frac{2 (2+n)}{2(1+n)^2} \tilde\Phi^2 + (\partial_{x_1}^2 + \partial_{x_2}^2)(\Phi + v \tilde \Phi) + 2 n \partial_u \tilde \Phi + \frac{1}{2}(\partial_{x^1}W_2 - \partial_{x^2}W_1)^2 - \partial_u(\partial_{x^1}W_1 + \partial_{x^2}W_2)\nonumber\\
    & - 2 (W_1\partial_{x_1} + W_2\partial_{x_2})\tilde\Phi - \tilde\Phi (\partial_{x^1}W_1 + \partial_{x^2}W_2)\label{eqPhi}\\
	0 =\mathcal{R}_{u1}&=\frac{1}{2}(\partial^2_{x^2}W_1 - \partial_{x^1}\partial_{x^2}W_2) + \frac{2+n}{2(1+n)} \partial_{x^1}\tilde \Phi\\
	0 =\mathcal{R}_{u2}&=\frac{1}{2}(\partial^2_{x^1}W_2 - \partial_{x^1}\partial_{x^2}W_1) + \frac{2+n}{2(1+n)} \partial_{x^2}\tilde \Phi\,.
\end{align}
The last two equations determine $\tilde \Phi$ in terms of $W_1$ and $W_2$, while $\Phi$ can subsequently be obtained from the Poisson-type Eq.~\eqref{eqPhi}. And, as consistency check, observe that for $n=0$ the equations become the Einstein vacuum equations of the VSI spacetimes \eqref{VSIsubclass}.In the $\tilde \Phi = 0$ case the Berwald Ricci tensor reduces to the Ricci tensor of the Lorentzian metric $g$, and hence the field equations become identical to Einstein's vacuum equations for the gyratonic $pp$-wave metric. Note that for $n=0$ we recover Einstein's equations corresponding to the VSI spacetimes \eqref{VSIsubclass}.

%++++++++++++++++++++++++++++++++++++++++++++++++%
\subsubsection{Non null-VGR Berwald spacetimes}\label{ssec:VGRCosmo}
Homogeneous and isotropic spacetimes are typically  encountered in cosmology. In Lorentzian geometry these \mbox{symmetry} demands on spacetime lead to a metric of Friedman-Lema\^itre-Robertson-Walker (FLRW) type. Finsler geometry has also been considered in the context of cosmological models \cite{Mavromatos:2010nk,Kouretsis:2013cga,Basilakos:2013hua,Papagiannopoulos:2017whb,Chang:2018msh,Hohmann:2016pyt}. In \cite{Pfeifer:2011xi} the most general homogeneous and isotropic Finsler spacetimes were derived. Applying this procedure to VGR spacetimes we find the Lagrangian 
\begin{align}\label{Lcosmo}
	L = \big(-(y^t)^2 + A^2(t) w^2 \big) \big(B(t)y^t\big)^n, \textrm{ with } w^2=\frac{(y^r)^2}{1-k r^2} + r^2 (y^\theta)^2 + r^2 \sin\theta^2 (y^\phi)^2\,
\end{align}
expressed in coordinates $(t,r,\theta,\phi;y^t,y^r,y^{\theta},y^{\phi})$. The Lorentzian metric $g$ is given by the FLRW metric, with $A(t)$ the scale factor and $k$ the spatial curvature constant. Evaluating condition \eqref{eq:Thm1cond} for the Lorentzian metric and $1$-form defining this VGR spacetime we find that it is of Berwald type if and only if:
\begin{align}
	B(t) = c\ A(t)^{-\frac{2+n}{n}}\,,
\end{align}
where $c$ is a constant. By direct calculation one can check that $B$ satisfies the VGR Berwald condition
\begin{equation}
\nabla_aB_b = -\frac{A'(t)}{n c A(t)^{-\frac{2}{n}}} \left(2(1+n)B_a B_b - n g(B,B) g_{ab}\right),
\end{equation}
with $C(x) = -(A'(t)/n c A(t)^{-2/n})$ and prime denoting derivative with respect to $t$. Note that  in this case $g(B,B)\neq 0$, and thus (\ref{Lcosmo}) is a non null-VGR Berwald spacetime. We compute next the geodesic spray employing Eq.~\eqref{eq:geodsprayvgrberwald}:
\begin{align}
	G^t &= - (y^t)^2 \frac{A'(t)}{A(t)}, \quad
	G^\theta = \frac{2}{r}y^r y^\theta - (y^\phi)^2 \cos\theta \sin\theta, \quad
	G^\phi = \frac{2}{r}y^r y^\phi + 2 \cot\theta\, y^\theta y^\phi,\\
	G^r &= \frac{k r}{1-kr^2}(y^r)^2 - r (1 - k r^2) ((y^\theta)^2 + \sin\theta^2 (y^\phi)^2)\,.
\end{align}
Note that the spray coefficients are independent of the parameter $n$ parametrizing the deviation from metric spacetime geometry, which is striking given the explicit dependence on $n$ of Eq.~\eqref{eq:geodsprayvgrberwald}. Thus the geometry of the homogeneous and isotropic VGR Berwald spacetime, i.e.\ the geodesic spray and its derived quantities, does not have a metric spacetime limit for $n\to0$, while the defining Lagrangian Eq.~\eqref{Lcosmo} does. This is a result of the specific form of function $C(x)$, which cancels the $n$-dependence of Eq.~\eqref{eq:geodsprayvgrberwald}. Seemingly, this type of geometry only exists in a proper ($n\neq 0$) Finsler setting. The deeper geometrical meaning and physical interpretation of this new family of spacetimes is unclear so far and will be investigated in future work.\\

The curvature scalar \eqref{eq:FCurvScal} can now be calculated from the geodesic spray, taking the surprisingly simple form:
\begin{align}\label{eq:homisoR}
	\mathcal{R} = 2 k \bigg(\frac{(y^r)^2}{1-k r^2} + r^2 (y^\theta)^2 + r^2 \sin\theta^2 (y^\phi)^2\bigg)\,.
\end{align}

For non-null VGR Berwald spacetimes the field equation do \emph{not} reduce to $\mathcal{R}_{ab}(x)=0$. and it is necessary to consider the full Eq.~\eqref{eq:FGravBerwald}, which in this case reads:
\begin{align}\label{eq:VGRBerwaldCosmo}
	- \frac{6 (y^t)^2}{A^2(t)} + w^2 \bigg(\frac{7n+5}{2 n^2 + n-1} + \frac{8 n (y^t)^2}{(2+n)(y^t)^2 + n w^2 A^2(t)}\bigg) = 0.
\end{align}
Here the scale factor $A(t)$ appears only through the Finsler Lagrangian and the L-metric. Performing the procedure outlined in section \ref{sec:BerwVGRDyn}, we can multiply the equation by the $y$-dependent denominator and obtain the tensorial equation \eqref{eq:VGRBerwaldTensorEq} on spacetime for this case, by applying four $y$-derivatives. The analysis of the resulting equation reveals that there exists no $A(t)$ such that \eqref{eq:VGRBerwaldCosmo} is solved. The only exception is the spatially flat case $k=0$; for such geometries any choice of $A(t)$ solves the field equation since, by equation \eqref{eq:homisoR}, $k=0$ implies $\mathcal{R}_{ab}(x)=0$ independently of $A(t)$.\\

Comparing homogeneous and isotropic VGR Berwald vacuum spacetimes to vacuum solutions of the Friedmann equations one finds quite some differences. In general relativity, for non-vanishing $k$, there exists the Milne Universe \cite{Carroll} as solution with $k=-1$, while for $k=0$ Minkowski spacetime is the only solution. For the generalizations of general relativity studied here we found that for $k\neq 0$ there exist no solutions, but for $k=0$ there exists a whole family of solutions parametrized by the functions $A(t)$. Note that the same is true for non action-based Finsler gravity equations such as $\mathcal{R} = 0$ \cite{Rutz} and $\mathcal{R}_{ab}=0$ \cite{Li:2012ty}. We would like to stress that these statements only hold for Berwald gravity vacuum dynamics; in the presence of matter there may exist interesting homogeneous and isotropic VGR Berwald spacetimes solving the field equations. The coupling between VGR spacetimes and matter will be investigated in future work.

%%%%%%%%%%%%%%%%%%%%%%%%%%%%%%%%%%%%%%%%%%%%%%%%%%%%%%%
\section{Discussion}
Finsler spacetimes are natural generalizations of Lorentzian spacetimes. They are viable candidates for an improved description and understanding of the geometry of spacetime, i.e. gravity \cite{Goenner:2008rr,Lammerzahl:2018lhw}. 
Finsler geometry immediately emerges in the context of modified dispersion relations. These may for example arise from an effective description of Planck scale quantum gravity effects \cite{Girelli:2006fw,Amelino-Camelia:2014rga,Letizia:2016lew}, or from field theories with field equations not (solely) defined by a Lorentzian  metric \cite{Kostelecky:2003fs,Punzi:2007di,Kostelecky:2010hs,Raetzel:2010je,Schreck:2015seb,Gurlebeck:2018nme}. Part of the motivation for this work lies in a particular instance of the latter, the very special relativity (VSR) framework by Cohen and Glashow \cite{Cohen:2006ky}. However, the variety of possible modified dispersion relations and corresponding Finsler geometries is vast, which complicates the investigation of the physical viability of general Finsler geometry as possible spacetime geometry. In order to derive physical observables and compare them with experiments it is therefore necessary to consider a specific class of Finsler spacetimes. This has been done for certain Finsler spacetimes at the solar system scale \cite{Lammerzahl:2012kw} and in a cosmological setting \cite{Hohmann:2016pyt}.\\

Berwald spacetimes are a particularly interesting class of Finsler spacetimes. They can be regarded as the \mbox{\emph{minimal}} Finsler generalization of Lorentzian spacetimes, i.e., the mildest deviation from metric geometry, and are physically relevant~\cite{Letizia:2016lew,Minguzzi2016}. We consider the Berwald generalization of Einstein's equations, derived from an action principle on the unit tangent bundle. We show that in this framework the vanishing of the Berwald Ricci tensor is a sufficient (although in general, not necessary) condition for the vacuum dynamics to be satisfied, which is analogous to the vanishing of the Ricci tensor in Einstein's vacuum equations.\\

Next we undertake a rigorous analysis of very general relativity (VGR)  spacetimes, which are curved \mbox{generalizations} of flat VSR spacetimes. We derive the geodesic spray, the fundamental building block of the corresponding geometry, and identify a necessary and sufficient condition for the spacetime to be of Berwald type (Theorem 1). We find necessary conditions for $(\mathcal{A},\mathcal{B})$-Finsler spacetimes, and in particular VGR spacetimes, to be not only Berwald but also have a geometry fully determined by the Levi-Civita connection of the Lorentzian metric involved (\mbox{Theorem 2}). This is closely related to similar results for Randers and Finsler $b$-spaces involving covariantly constant $1$-forms~\cite{matsumoto1974,Hashiguchi1975,Shibata1977,Kikuchi1979,Kostelecky:2011qz}. Moreover, we show that the Berwald VGR field equations are equivalent to the vanishing of the Berwald Ricci tensor if and only if the $1$-form appearing in the VGR line element is null with respect to the Lorentzian metric (Theorem 3). Hence in general the vanishing of the Finsler Ricci tensor, previously proposed in the literature as \mbox{vacuum} field equation, is not equivalent to the unit tangent bundle action-based Finsler gravity field equation. \\

Finally, we prove that the Berwald VGR class of spacetimes is non-empty by presenting two novel examples, namely a Finslerian generalization of VSI spacetimes in Einstein's gravity, and the most general homogeneous and isotropic VGR spacetime. We also derive the corresponding vacuum field equations, which for (a subclass of) the first example adopt the same form as the one in Einstein's gravity. In the second example it turns out that all homogeneous and isotropic VGR Berwald spacetimes which are spatially flat solve the vacuum equations, while there exists no solution for spatially non-flat geometries (which occurs in Einstein's gravity).\\

A next step in the analysis of VGR Berwald spacetimes would be to perform an exhaustive classification of all possible such spacetimes, by studying the most general form of the Lorentzian metric and $1$-form in the line element which are compatible with the Berwald requirement. A complete mathematical classification of general $(\mathcal{A},\mathcal{B})$-Finsler spacetimes may also be done in the future, based on their geodesic spray presented in the appendix.

A physically interesting research direction is the addition of matter to the investigated Berwald vacuum dynamics. This coupling could be realized in different ways, for example by formulating the VSR framework of Cohen and Glashow in the same language as the Finslerian gravitational action \cite{Pfeifer:2011xi}, or by stating field theories directly on the tangent bundle where they couple naturally to a Finslerian geometry \cite{Pfeifer:2011tk}. A perfect fluid coupling to VGR spacetimes, particularly relevant as source for the homogeneous and isotropic case and its application to cosmology, may be achieved by employing the kinetic gases description on the tangent bundle as introduced by Ehlers \cite{Ehlers2011} and later connected to Finsler geometry in \cite{Hohmann:2015ywa}.

%%%%%%%%%%%%%%%%%%%%%%%%%%%%%%%%%%%%%%%%%%%%%%%%%%%%%%%
\begin{acknowledgments}
	The authors would like to thank E.\,Minguzzi and J.D.\,Barrow for their helpful comments and remarks. A.\,Fuster would like to acknowledge A.\,Ach\'{u}carro for everything. The work of A.\,Fuster is part of the research program of the Foundation for Fundamental Research on Matter (FOM), which is financially supported by the Netherlands \mbox{Organisation} for Scientific Research (NWO). C.\,Pfeifer gratefully acknowledges support of the European Regional Development Fund through the Center of Excellence TK133 ``The Dark Side of the Universe''.
\end{acknowledgments}

\appendix

%%%%%%%%%%%%%%%%%%%%%%%%%%%%%%%%%%%%%%%%%%%%%%%%%%%%%%%
\section{The Landsberg tensor and the Berwald covariant derivative}\label{app:def}
In the formulation of the Finsler spacetime vacuum dynamics \eqref{eq:Fgrav} we encountered the Berwald covariant derivative and the Landsberg tensor, which we will properly introduce here.

The Berwald linear covariant derivative $\nabla^B$ is a linear covariant derivative on the tangent bundle of a Finsler spacetime $(M,L)$ defined by the following action on the horizontal-vertical basis of the tangent spaces of the tangent bundle
\begin{align}
	\nabla^B_{\delta_a}\delta_b &= \bar{\partial}_a N^{q}{}_{b}\delta_q,\quad \nabla^B_{\delta_a}\bar{\partial}_b = \bar{\partial}_a N^{q}{}_{b}\bar{\partial}_q,\\
	\nabla^B_{\bar{\partial}_a}\delta_b &= 0,\qquad \qquad \nabla^B_{\bar{\partial}_a}\bar{\partial}_b = 0\,,
\end{align}
where $N^{a}{}_{b}$ are the connection coefficients of the Cartan non-linear connection \eqref{eq:CNL} and $\delta_a$ are the horizontal basis \eqref{eq:horr}.

The Landsberg tensor $S$ is the difference between the $\delta$-Christoffel symbols
\begin{align}
	\Gamma^{\delta a}{}_{bc} = \frac{1}{2}g^{Laq}\big(\delta_b g^L_{cq} + \delta_c g^L_{bq} - \delta_q g^L_{bc} \big)
\end{align}
and the connection coefficients of the Berwald connection \cite{Pfeifer:2011tk}
\begin{align}\label{eq:Stensor}
	S^a{}_{bc} = \Gamma^{\delta a}{}_{bc} - \bar{\partial}_b N^{a}{}_{c}\,.
\end{align}

%%%%%%%%%%%%%%%%%%%%%%%%%%%%%%%%%%%%%%%%%%%%%%%%%%%%%%%
\section{The geodesic spray of $(\mathcal{A},\mathcal{B})$-Finsler spacetimes}\label{app:geodSprayL(A,B)}
To derive the geodesic spray of a VGR spacetime in section \ref{ssec:VGRL} and to proof Theorem 2 of section \ref{ssec:BerwVGR}, it is most convenient to consider general $(\mathcal{A},\mathcal{B})$-Finsler spacetimes and in particular the VGR Finsler Lagrangian expressed as
\begin{align}
	L = \mathcal{A}\mathcal{B}^n\,,
\end{align}
with variables $\mathcal{A} = g(y,y) = g_{ab}(x)y^ay^b$ and $\mathcal{B} = B(y) = B_a(x)y^a$.

This can be done for general $(\mathcal{A},\mathcal{B})$-Finsler spacetimes based on Lagrangians of the form $L = L(\mathcal{A},\mathcal{B})$ as follows. We need to calculate $\partial_c L$, $y^d\partial_d\bar{\partial}_cL$ and the inverse Finsler metric $g^{L ab}$, as displayed in Eq.~\eqref{eq:geodsprayL}.

Let us write down the results and use the following abbreviations: $\partial_\mathcal{A} = \frac{\partial}{\partial \mathcal{A}}$, $\partial_\mathcal{B} = \frac{\partial}{\partial \mathcal{B}}$, $\Gamma^a{}_{bc}$ are the Christoffel symbols of the metric $g$ and as short-hand notation we write $g(B,B) = g^{ab}B_a B_b$, where we identified the metric dual $g^{-1}(B, \cdot )$ with $B$ itself for the sake of readability. The calculations below were performed with help of the computer algebra program XAct for Mathematica.

\begin{itemize}
	\item The first derivatives of $L$
	\begin{align}
	\partial_cL &= \partial_{\mathcal{A}}L\ \partial_c g_{ab} y^ay^b + \partial_{\mathcal{B}}L\ \partial_cB_a y^a\\
	\bar\partial_cL &= 2 \partial_{\mathcal{A}}L\  g_{ac} y^a + \partial_{\mathcal{B}}L\ B_c\,.
	\end{align}
	\item The mixed derivative of $L$
	\begin{align}
	y^d \partial_d\bar{\partial}_cL
	&= 2 \partial_{\mathcal{A}}L\ y^my^d (g_{mi}\Gamma^i{}_{dc} + g_{ca}\Gamma^a{}_{dm}) + \partial_{\mathcal{B}}L\ y^d(\nabla_d B_c + \Gamma^{m}{}_{dc}B_m)\\
	&+ \partial_{\mathcal{A}}\partial_{\mathcal{A}}L\ 4 y_c y^d\Gamma^m{}_{db}y_my^b + \partial_{\mathcal{B}}\partial_{\mathcal{B}}L\ B_c y^b y^d(\nabla_d B_b + \Gamma^m{}_{db} B_m)\nonumber\\
	&+\partial_{\mathcal{B}}\partial_{\mathcal{A}}L\ (2 y_c y^d (\nabla_d B_b + \Gamma^m{}_{db} B_m) y^b + B_c 2 y_m y^d \Gamma^m{}_{db}y^b)\,.\nonumber
	\end{align}
	\item The $L$-metric
	\begin{align}
	g^L_{ab} = g_{ab} \partial_{\mathcal{A}}L + 2 y_a y_b \partial_{\mathcal{A}}\partial_{\mathcal{A}}L + (y_a B_b + y_b B_a)\partial_{\mathcal{A}}\partial_{\mathcal{B}}L 
	+ \frac{1}{2} B_a B_b\partial_{\mathcal{A}}\partial_{\mathcal{A}}L \,.
	\end{align}
	\item The inverse $L$-metric
	\begin{align}
	g^{Lab} &= \frac{1}{\partial_{\mathcal{A}}L}g^{ab} \\
	&+ \frac{1}{Q} \big(B^a y^b + B^b y^a\big) \big(\mathcal{B}\ \partial_{\mathcal{A}}\partial_{\mathcal{A}}L\ \partial_{\mathcal{B}}\partial_{\mathcal{B}}L  - \partial_{\mathcal{A}}\partial_{\mathcal{B}}L (\mathcal{B}\ \partial_{\mathcal{A}}\partial_{\mathcal{B}}L + \partial_{\mathcal{A}}L)\big)\nonumber\\
	&+ \frac{1}{Q} y^a y^b \big( g(B,B)\ (\partial_{\mathcal{A}}\partial_{\mathcal{B}}L)^2 - \partial_{\mathcal{A}}\partial_{\mathcal{A}}L ( g(B,B)\  \partial_{\mathcal{B}}\partial_{\mathcal{B}}L + 2 \partial_{\mathcal{A}}L) \big)\nonumber\\
	&+ \frac{1}{Q} B^a B^b \big( \mathcal{A}\ (\partial_{\mathcal{A}}\partial_{\mathcal{B}}L)^2\ - \tfrac{1}{2}\partial_{\mathcal{B}}\partial_{\mathcal{B}}L (2 \mathcal{A}\ \partial_{\mathcal{A}}\partial_{\mathcal{A}}L + \partial_{\mathcal{A}}L)\big)\,,\nonumber
	\end{align}
	where $Q$ is given by
	\begin{align}
	Q = \partial_{\mathcal{A}}L\bigg( (\partial_{\mathcal{A}}L)^2 
	&+ \partial_{\mathcal{A}}L \big(\tfrac{1}{2} g(B,B)\ \partial_{\mathcal{B}}\partial_{\mathcal{B}}L + 2 \mathcal{B}\ \partial_{\mathcal{A}}\partial_{\mathcal{B}}L + 2 \mathcal{A}\ \partial_{\mathcal{A}}\partial_{\mathcal{A}}L\big)\\
	&+ (\mathcal{B}^2 - g(B,B)\ \mathcal{A})\big(  (\partial_{\mathcal{A}}\partial_{\mathcal{B}}L)^2 - \partial_{\mathcal{A}}\partial_{\mathcal{A}}L\ \partial_{\mathcal{B}}\partial_{\mathcal{B}}L \big)\bigg)\,.\nonumber
	\end{align}
\end{itemize}
These ingredients can be combined into Eq.~\eqref{eq:geodsprayL} and we obtain
\begin{align}\label{eq:Ga}
 G^a(x,y) &= \Gamma^{a}{}_{bc}(x)y^by^c \nonumber \\
&+ \frac{1}{2}\frac{\partial_{\mathcal{B}}L}{\partial_{\mathcal{A}}L} y^b g^{ac} (\nabla_b B_c - \nabla_c B_b) \nonumber\\
&+ \frac{1}{4}\frac{1}{Q} \bigg( y^b B^c (\nabla_c B_b - \nabla_b B_c)\ \partial_{\mathcal{B}}L  + y^b y^c (\nabla_c B_b + \nabla_b B_c) \ \partial_{\mathcal{A}}L\bigg) \nonumber\\
&\times\bigg[ 2 y^a \big( \partial_{\mathcal{A}}\partial_{\mathcal{B}}L\ ( \partial_{\mathcal{A}}L + \mathcal{B}\ \partial_{\mathcal{A}}\partial_{\mathcal{B}}L ) - \mathcal{B}\ \partial_{\mathcal{A}}\partial_{\mathcal{A}}L\ \partial_{\mathcal{B}}\partial_{\mathcal{B}}L \big) \nonumber\\
&+ B^a \big(  \partial_{\mathcal{B}}\partial_{\mathcal{B}}L\ ( \partial_{\mathcal{A}}L\ + 2 \mathcal{A}\ \partial_{\mathcal{A}}\partial_{\mathcal{A}}L) - 2 \mathcal{A}\ (\partial_{\mathcal{A}}\partial_{\mathcal{B}}L)^2\big) \bigg] \,.
\end{align}
Note that the corresponding expression of the geodesics spray for $(\alpha,\beta)$-Finsler spaces (the positive definite predecessor of $(\mathcal{A},\mathcal{B})$-Finsler spacetimes) has been obtained for example in \cite{Matsumoto} and in \cite{ShenBacsoCheng}. Since the homogeneity of $L$ depends on the Finsler spacetime one considers, and since we seek to discuss general $(\mathcal{A},\mathcal{B})$-Finsler spacetimes we did not introduce a zero-homogeneous variable $s = \frac{\mathcal{B}^2}{\mathcal{A}}$ as done in the discussions on $(\alpha,\beta)$-Finsler spaces, but kept $\mathcal{A}$ and $\mathcal{B}$ separately as variables. 

The expression derived so far lead to an easy proof of Theorem 2, which we repeat here again for easy readability of this appendix.\\

\noindent \textbf{Theorem 2.} \emph{Let $(M,L)$ be a $(\mathcal{A},\mathcal{B})$-Finsler spacetime, i.e. $L=L(\mathcal{A},\mathcal{B})$ with $\mathcal{A}=g(y,y)$ and $\mathcal{B}=B(y)$. A \mbox{sufficient} condition for $(M,L)$ to be a Berwald spacetime is that $B$ is covariantly constant with respect to the Lorentzian \mbox{metric}~$g$. The geodesic spray of $(M,L)$ is then given by $G^a = \Gamma^a{}_{bc}y^by^c$, where $\Gamma^a{}_{bc}$ are the Christoffel symbols of the metric.}\\

\noindent \textbf{Proof:}\\
For $\nabla_aB_b=0$ it is clear that \eqref{eq:Ga} becomes quadratic in the velocities. It is given by the Christoffel symbols of the metric defining the $(\mathcal{A},\mathcal{B})$-Finsler spacetime
\begin{align}\label{eq:propA1}
G^a = \Gamma^{a}{}_{bc}(x)y^by^c\,.\quad \square
\end{align}
Due to the complicated structure of the geodesic spray, like  the fact that Randers spaces are Berwald spaces if and only if the $1$-form $B$ is covariantly constant with respect to $g$ \cite{Bao}, stronger statements cannot be formulated in general. Only for specific examples such statements are possible. In this article we establish such statements for VGR spacetimes of the form $L = \mathcal{A}\mathcal{B}^n$ in section \ref{ssec:BerwVGR}.

%%%%%%%%%%%%%%%%%%%%%%%%%%%%%%%%%%%%%%%%%%%%%%%%%%%%%%%
\bibliographystyle{utphys}
\bibliography{VGRBerwald}

\end{document}